\documentclass[aps,prx,twocolumn, superscriptaddress]{revtex4-2}

\usepackage{float}
\usepackage{graphicx}
\usepackage{physics}
\usepackage{amssymb}
\usepackage{amsmath}
\usepackage{color}
\usepackage{graphicx}
\usepackage{dcolumn}
\usepackage{epstopdf}
\usepackage{bm}
\usepackage{natbib}
\usepackage{tikz}
\usepackage{siunitx}
\usepackage{lineno}
\usepackage{soul}
\DeclareSIUnit{\dBm}{\deci\belmilliwatt}
\usepackage[colorlinks=true,citecolor=blue]{hyperref}

\newcommand{\orcid}[1]{\href{https://orcid.org/#1}{\includegraphics[width=8pt]{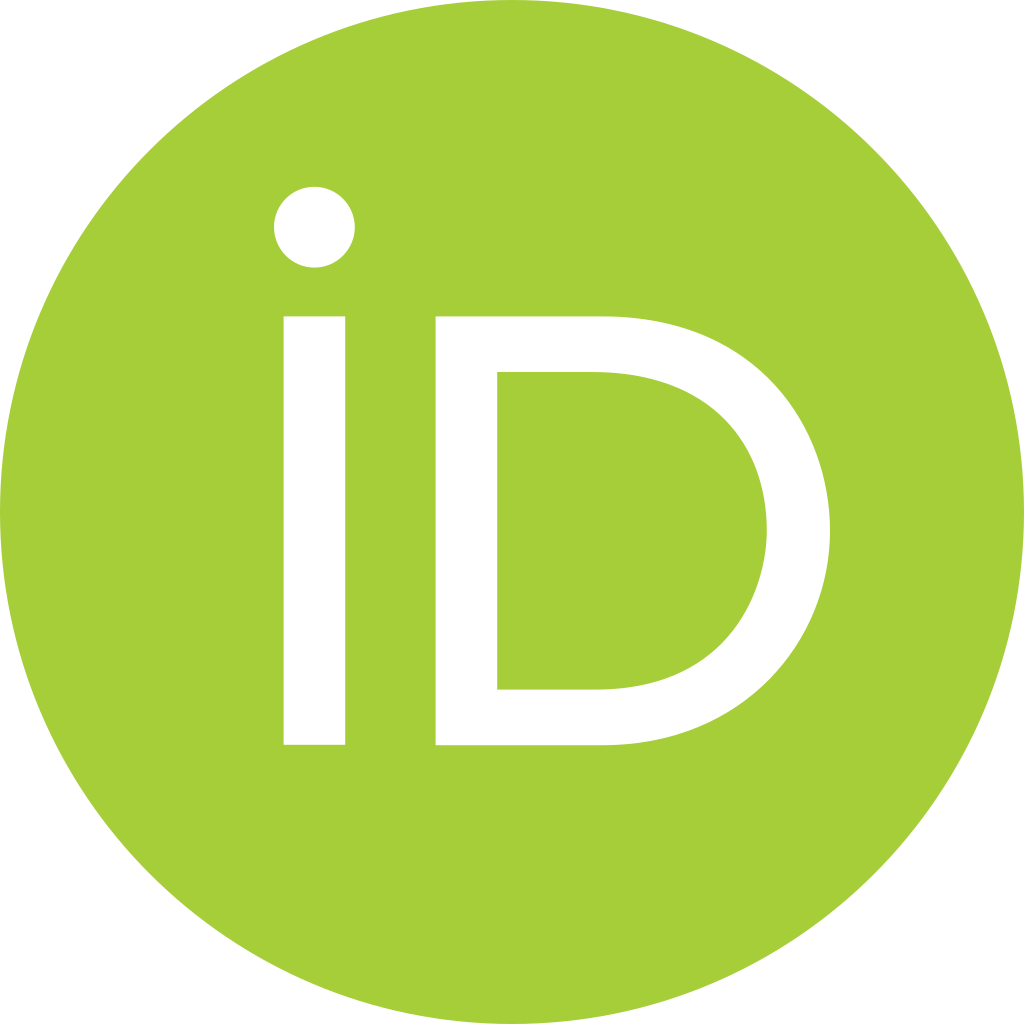}}}

\begin{document}
\title{Near-Unity Charge Readout Signal in a Nonlinear Resonator without Matching the Sensor Dissipation}

\author{Harald Havir \orcid{0000-0002-2944-986X}}
\affiliation{NanoLund and Solid State Physics, Lund University, Box 118, 22100 Lund, Sweden}

\author{Andrea Cicovic}
\affiliation{NanoLund and Solid State Physics, Lund University, Box 118, 22100 Lund, Sweden}

\author{Pierre Glidic}
\affiliation{NanoLund and Solid State Physics, Lund University, Box 118, 22100 Lund, Sweden}
\affiliation{Université Lyon 1, CNRS, Institut Lumière Matière, UMR5306, F-69100, Villeurbanne, France}
\author{Subhomoy Haldar}
\affiliation{NanoLund and Solid State Physics, Lund University, Box 118, 22100 Lund, Sweden}
\affiliation{Department of Physics, Indian Institute of Technology Kanpur, Uttar Pradesh 208016, India}

\author{Sebastian Lehmann}
\affiliation{NanoLund and Solid State Physics, Lund University, Box 118, 22100 Lund, Sweden}

\author{Kimberly A. Dick}
\affiliation{NanoLund and Solid State Physics, Lund University, Box 118, 22100 Lund, Sweden}
 \affiliation{Center for Analysis and Synthesis, Lund University, Box 124, 22100 Lund, Sweden}
\author{Ville F. Maisi \orcid{0000-0003-4723-7091}}
\email{ville.maisi@ftf.lth.se}
\affiliation{NanoLund and Solid State Physics, Lund University, Box 118, 22100 Lund, Sweden}

\begin{abstract}
    In this paper, we present a nonlinear resonator performing the readout of a double quantum dot charge state via a charge-sensing quantum dot. We show that by driving the resonator in the nonlinear regime, we achieve a near-unity signal for a dissipative sensor. This despite not satisfying the sensor impedance matching requirements necessary for such large signals in the linear regime.
    Our experiments, supported by numerical calculations, demonstrate that the signal increase stems from the sensor dissipation shifting the onset of the nonlinear resonator response.
    By lifting the matching requirement, we open up an avenue to ultra-fast charge detectors as the resonator input-output coupling - setting the detector bandwidth - does not have to match to the typically much slower sensor dissipation rate.
\end{abstract}
 
\date{\today}
\maketitle

\section{Introduction}
Resolving the position of a single electron fast~\cite{Schoelkopf1998, Pekola2013, Vigneau2023} is a cornerstone in mesoscopic physics. It enables for example reading out spin qubits~\cite{Mi2018, Zheng2019, Chatterjee2021, Oakes2023, Burkard2023}, highly accurate metrology standards \cite{Keller1999, Gustavsson2009}, and probing electron transport statistics \cite{Albert2011, Kung2012, Wagner2017, Tanttu2019}. The fastest charge detection is currently achieved with dissipative sensors reaching near the quantum~\cite{Schoelkopf1998} or shot-noise limit~\cite{Keith2019}.
One of the key requirements for the fast dissipative sensors is that one uses a resonator circuit to match the sensor dissipation rate $\kappa_\text{s}$ to the coupling rate $\kappa_\text{c}$ between the readout line and the resonator circuit~\cite{Turin2003, Ares2016, Vigneau2023}.

Fulfilling this matching condition, $\kappa_c = \kappa_s$, is equivalent of the usual impedance matching where the combined impedance of the sensor resistance and a lossless resonator circuit matches to the readout line impedance. This yields a unity change $|\Delta r| = 1$, to the measured reflection coefficient $r$ between the cases with and without the sensor dissipation $\kappa_s$. If the resonator has any other spurious losses with a rate $\kappa_i$, the response $|\Delta r|$ gets suppressed. Therefore the internal losses need to be low, $\kappa_\text{c} \gg \kappa_\text{i}$, to maintain the near unity change.

A detector operating under these conditions $\kappa_\text{s} = \kappa_\text{c} \gg \kappa_\text{i}$ thus achieves the unity readout signal enabled by impedance matching.
Increasing the detection speed determined by the resonator linewidth requires therefore that both $\kappa_\text{s}$ and $\kappa_\text{c}$ are increased. Otherwise the readout signal is suppressed, which slows down the detection speed as longer averaging time is needed to reach sufficient signal-to-noise ratio.

In this paper, we show that the matching requirement can be circumvented.
We use a nonlinear resonator to turn the dissipative response to a frequency shift, allowing us to use this shift similarly as done for inductive superconducting devices~\cite{Sillanpaa2004, Thyagarajan2024}, dispersive QD sensors~\cite{Zheng2019, Ares2016, Oakes2023, Vigneau2023}, and superconducting qubits~\cite{Wallraff2005, Lupa2006} that all have a frequency shifting response inherently.
This results in a near-unity reflection response for the dissipative response despite not following the above matching requirements.
We demonstrate further experimentally that the charge readout speed is increased by one order of magnitude in the nonlinear case, as compared to the corresponding linear case, and analyze that $\kappa_\text{s}$ is not directly limiting the maximum attainable speed as is the case for the linear detection. 
Furthermore, the nonlinear operation mode allows for a protection scheme where two charge states are protected from the readout signal: for one of the states, the sensor is not conducting, achieving a protection the same way as spin qubit readout in Refs.~\cite{Morello2010, Park2025}, and for the other, the nonlinearity shifts the resonator mode away from the frequency of the readout signal, hindering the readout signal to drive the resonator efficiently. 
The nonlinearities therefore enable both a higher readout bandwidth as well as new device concepts (the protection scheme and a way to circumvent the matching requirement and yet obtain a near-unity signal) for the dissipative sensors.

\section{Results}

\begin{figure*}[t]
    \centering
    \includegraphics[width=\textwidth]{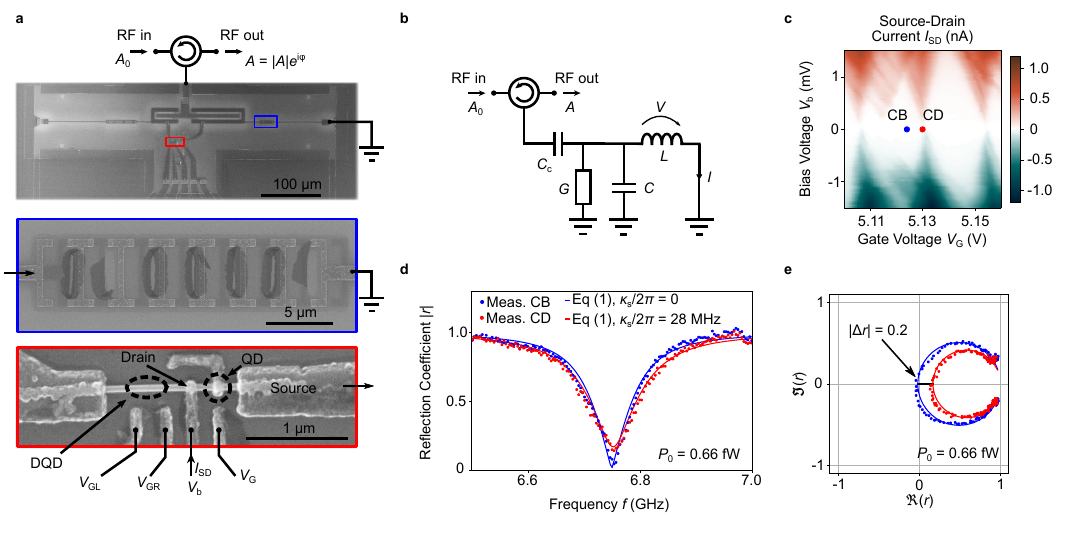}
    \caption{\textbf{Device and its response in the linear regime.}
    \textbf{a} Scanning electron micrographs of the measured device.
    The input/output port for RF signals at the top of the figure couples to the resonator via the longer c-shaped finger capacitor on the right. 
    The inductance for the resonator is achieved with an array of $N=13$ SQUIDs (blue) in series. 
    An InAs nanowire with crystal-phase defined barriers contains the sensor QD and a DQD, with metallic connections defining the source- drain- and gate contacts to the QD and DQD.
    \textbf{b} A circuit model of the device. 
    The input/output port couples to the resonator via a capacitance $C_\text{c}$, which defines the resonance frequency together with the capacitance $C$ to ground and inductance $L$ of the SQUID-array. 
    The QD adds a conductance $G$. 
    \textbf{c} Measured current $I_\text{SD}$ through the quantum dot as a function of the gate voltage $V_\mathrm{G}$ and bias voltage $V_\mathrm{b}$. 
    An additional Coulomb diamond with excited level structure is also visible in the response. This is transport from the DQD which connects to the same dc measurement line.
    \textbf{d} Reflection coefficient $|r|$ as a function of the drive frequency $f$ in the linear regime at Coulomb blockade (CB, blue) and at Coulomb degeneracy (CD, red) with input power $P_0 = 0.66$ fW. 
    The solid lines are fits to Eq. \eqref{eq:reflAmp}.
    \textbf{e} The data of panel \textbf{d} plotted in the IQ-plane. 
    The maximum difference in measured amplitude between the CB and CD cases is $0.2A_0$.} 
    \label{fig:dev}
\end{figure*}

The measured device, presented in Fig.~\ref{fig:dev}~a, contains a QD-based charge sensor (red) in the same geometry as Refs. \cite{Haldar2025} and a nonlinear resonator. 
The nonlinearities in the resonator are achieved with an array of $N=13$ superconducting quantum interference devices (SQUIDs) in series (blue)~\cite{Hermon1996, Castellanos-Beltran2007, Bell2012, Altimiras2013, Stockklauser2017, Scarlino2019}. 
As shown in the circuit diagram of Fig.~\ref{fig:dev}~(b), this array yields an inductance $L$ of the resonator that connects on one end to ground, while the other end is floating. 
The floating side therefore has an oscillating voltage $V$ of the resonance mode. 
An in-going radio frequency (RF) readout signal with amplitude $A_0$ couples to this voltage via a finger capacitor $C_\text{c}$. 
The source contact of the sensor QD also connects directly to the voltage. 
As the drain side of the QD is shunted capacitively to ground, the QD transport conductance $G$ adds dissipation directly into the resonator \cite{Harabula2017, Havir2023, Vigneau2023}.

Figure~\ref{fig:dev}~(c) presents the electrical transport of the sensor QD, measured in a dilution refrigerator with electronic base temperature $T =$ \SI{50}{\milli \kelvin}.
Here we measure the electrical current $I_\text{SD}$ as a function of DC bias voltage $V_\mathrm{b}$ and gate voltage $V_\mathrm{G}$.
We see that at a gate voltage $V_\mathrm{G} = 5.13$ V, the QD is conducting current with conductance $G = $ \SI{0.4}{\micro \siemens} at the Coulomb degeneracy (CD) while at larger and smaller $V_\mathrm{G}$ the conduction stops due to the Coulomb blockade (CB).
Figures~\ref{fig:dev}~(d) and (e) present the corresponding resonator response, showing the measured resonator reflection coefficient $|r| = |A/A_0|$ as a function of the input drive frequency $f$ in the linear regime at input power $P_0 = 0.66$~fW.
We see that making the QD conducting increases the linewidth of the response, along with a change in the reflection coefficient $\left|\Delta r\right| = 0.2$ at the resonance frequency $\omega_\text{r}/2\pi = 6.75$ GHz.

To determine the coupling $\kappa_\text{c}$ and losses $\kappa_\text{i}$ and $\kappa_\text{s}$ of our device, we fit the reflection response~\cite{Goppl2008, Probst2015, Havir2023, Rieger2023} of the form
\begin{equation}
    r = A/A_0 = 1 - \frac{\kappa_\text{c}}{\kappa/2 - i(\omega - \omega_\text{r})},
    \label{eq:reflAmp}
\end{equation}
where $\kappa = \kappa_\text{c} + \kappa_\text{i} + \kappa_\text{s}$ is the linewidth of the resonator.
From the CB data, we obtain the bare resonator parameter values $\kappa_\text{c}/2\pi = 62$ MHz and $\kappa_\text{i}/2\pi = 60$ MHz with $\kappa_\text{s} = 0$. 
The CD case is then fitted with $\kappa_\text{s}/2\pi = 28$ MHz as the only free parameter. 
Adding only this dissipation explains the changes in response from CB to CD.
Therefore the QD conductance adds dissipation to the resonator.
As our resonator with $\kappa_\text{i} \approx \kappa_\text{c} > \kappa_\text{s}$ is not fulfilling the matching conditions, the sensor QD changes the response only by $\left|\Delta r\right| = 0.2$.

{
\begin{figure*}[hbt!]
    \centering
    \includegraphics[width=1.0\textwidth]{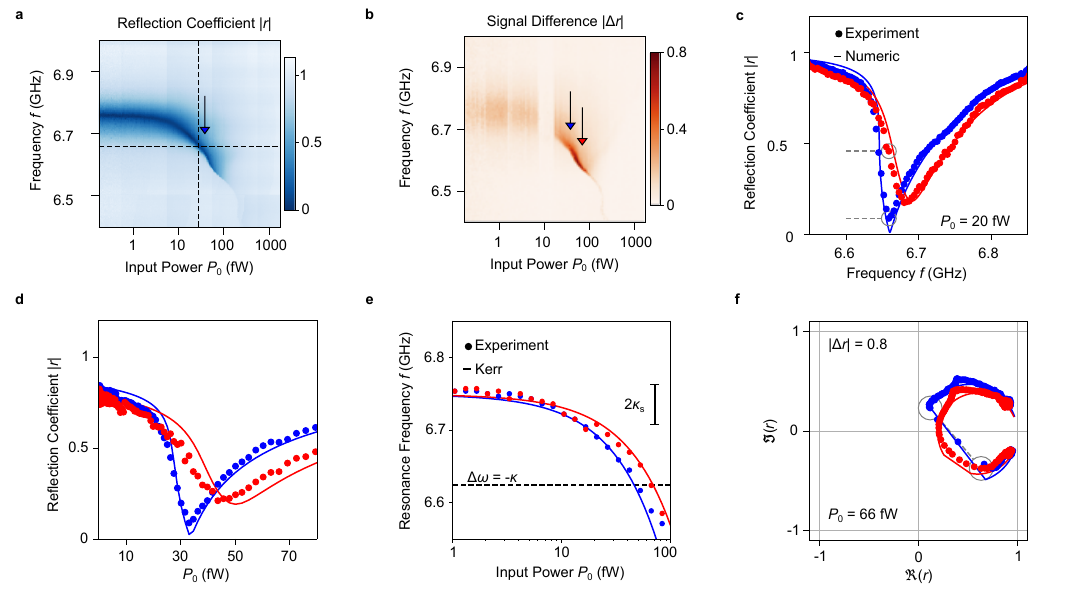}
    \caption{
    \textbf{The response in the nonlinear regime.}
    \textbf{a} The reflection coefficient $|r|$ as a function of drive frequency $f$ and input power $P_0$ for CB. 
    \textbf{b} The reflection coefficient change $|\Delta r|$ between CB and CD.
    The blue and red arrow in panels \textbf{a} and \textbf{b} indicate the threshold powers of Eq.~(\ref{eq:Pth}) in CB, $P_0 = 43$ fW, and CD, $P_0 = 84$ fW, respectively.
    \textbf{c} \& \textbf{d} Frequency \& power  response along the vertical/horizontal dashed lines of panel \textbf{a}, $P_0 = 20$ fW / $f = 6.659$ GHz for CB (blue) and CD (red). 
    The lines are numerical calculations made with the circuit model presented in the Supplemental Materials Section C, and using the low power response parameter values as input parameters.
    \textbf{e} The resonance frequency $f$ as extracted from the minimum amplitude response of the experimental data (dots) and as predicted by the Kerr shift (lines). 
    The solid bar shows the predicted frequency shift of $\Delta\omega = 2\kappa_\text{s}$.
    \textbf{f} The resonator response with $P_0 = $ \SI{66}{\femto \watt} shown in the IQ plane.
    A signal of $|\Delta r|= 0.8$ is found between the CB and CD cases at frequency $f = 6.593$ GHz, indicated by the open circles and the dashed line highlighting the highest $|\Delta r|$ case with fixed $P_0$ and $f$.
    }
    
    \label{fig:nonlinChar}
\end{figure*}

Next we consider the nonlinear regime of the resonator.
For this, Fig.~\ref{fig:nonlinChar}a repeats the measurement of Fig.~\ref{fig:dev}d for different input powers $P_0$.
At $P_0 < 1$  fW, the response is that of Fig.~\ref{fig:dev} (d) and does not depend on $P_0$.
This is the linear response regime.
At $P_0 > 1$ fW the resonator mode shifts to lower frequencies, and eventually vanishes.
This is due to the cosine-shaped potential of the Josephson junctions, which leads to nonlinear inductance at a large input power \cite{Castellanos-Beltran2007, Tholen2007, Bishop2010, Bell2012}.
The signal difference $\left|\Delta r\right|$ between the CB and CD cases is shown in Fig.~\ref{fig:nonlinChar}b.
Here, we see again that $\left|\Delta r\right| = 0.2$ in the linear response regime, which increases to the near-unity value of $\left|\Delta r\right| = 0.8$ in the nonlinear regime.

Figures \ref{fig:nonlinChar}c-d show line cuts along the dashed lines of Fig.~\ref{fig:nonlinChar}a for CB and CD.
Here, the response is no longer Lorentzian, but displays a steep slope on the low-frequency side arising from bifurcation in the nonlinear resonator \cite{Bishop2010, Ong2011, Frasca2023}.
Furthermore, making the QD conducting shifts the response in frequency.
This is in stark contrast to the linear response of Fig.~\ref{fig:dev} (d) where the shift is vanishingly small.
As the frequency shift moves the resonator response, the readout signal between the two states is much larger as indicated by the dashed lines in Fig.~\ref{fig:nonlinChar} (c), despite not satisfying the matching requirement.

To model the response theoretically, we calculate the reflection coefficient $r$ of the circuit in Fig.~\ref{fig:dev}b.
We do this using the semiclassical Josephson equations $I = I_0 \sin{\phi}$ and $V = \frac{\hbar N}{2e}\frac{\partial\phi}{\partial t}$,
for the current $I$ through, and voltage $V$ across the JJ array, with $\phi$ the difference in the superconducting phase across a single junction. 
We solve $\phi(t)$ using a harmonic balance method up to the first fundamental mode of the resonator for a given sinusoidal input voltage drive, with the approximation that $\sin(\phi) = \phi - \phi^3/6$, see supplemental material for details.
The result of the numerical calculations are shown as the solid lines in Fig.~\ref{fig:nonlinChar} (c, d) with the above parameter values determined in the linear response regime together with the resistance of a single SQUID $R_\mathrm{J} = $ \SI{1.5}{\kilo \ohm}.
This resistance was measured at room temperature to $R = $ \SI{1.25}{\kilo \ohm}, and adjusted to account for an $\approx 20\ \%$ increase at low temperatures \cite{Holmqvist2008}.
Together with the known number of junctions in the array as a final input parameter, this leaves no free variables in the numerical calculations apart from a 20 \% (0.8 dB) correction to the input power.
This correction was made to the RF calibration based on the onset of the non-linearities, and is within the typical uncertainty of $\sim$1 dB of the microwave setups.
As $\kappa_\text{s}$ is the only difference between the CB and CD calculation, the theory model confirms that the nonlinear resonator turns the small dissipative response of the QD in the linear regime to a much stronger response via the induced frequency shift in Fig. \ref{fig:nonlinChar} (c).

We can further quantify the onset of the nonlinear regime and the frequency shift analytically by considering the Kerr term~\cite{Koch2007} of the resonator $E_\mathrm{K} = -E_\text{C}/N^2$.
This term leads to a resonance frequency shift
\begin{equation}
    \frac{\omega_\text{K}}{\omega_\text{r}}  =
    \frac{E_K n}{\hbar\omega_r} =
    -\frac{E_\text{C}}{\hbar\omega_\text{r} N^2}n = -\frac{4\pi Z_\text{r} \kappa_\text{c}}{R_\text{Q} N^2\kappa^2}\frac{P_0}{\hbar \omega_\text{r}},
    \label{eq:nonLinShift}
\end{equation}
where $E_\text{C} = e^2/2C_\mathrm{r}$ is the charging energy of the resonator island, $R_\text{Q} = h/e^2$ the resistance quantum, $n = (4\kappa_\text{c}P_0)/(\hbar\omega_r\kappa^2)$ the number of photons in the microwave cavity \cite{Andersson2025}, $Z_\text{r} = \sqrt{L/C_\mathrm{r}}$ the characteristic impedance of the resonator and $N$ the number of junctions/SQUIDs in the array.
In Fig. \ref{fig:nonlinChar} (e) we plot the predicted resonator frequency $\omega = \omega_r + \omega_\mathrm{K}$, and find a good agreement with the experimental data.
Here the resonance frequency of the experimental data is determined from the minimum of the reflected signal.

To quantify the sensor response, we consider the resonator frequency change $\Delta\omega_\mathrm{K}$ that results from the sensor loss $\kappa_\text{s}$.
From Eq. \eqref{eq:nonLinShift}, the resulting frequency change is 
\begin{equation}
    \Delta\omega_\text{K}/\omega_\text{r} = \frac
    {4\pi Z_\mathrm{r}\kappa_\text{c}}
    {R_Q N^2 \kappa_\text{CB}^2}
    \frac{P_0}{\hbar\omega_r}
    \left(1 - \frac
    {\kappa_\text{CB}^2}
    {(\kappa_\text{CB}+\kappa_\text{s})^2}
    \right),
    \label{eq:dissipationShift}
\end{equation}
where $\kappa_\text{CB} = \kappa_\text{c} + \kappa_\text{i}$.
To simplify the consideration further, we assume the sensor to contribute only a small addition to the resonator dissipation $\kappa_\text{s} \ll \kappa_\text{CB}$.
Also, we consider the non-linear response regime at $\omega_\text{K} = -\kappa$.
Under this condition, the frequency shift is comparable to the linewidth, making it prominent in the response, while just reaching the bifurcation regime~\cite{Ong2011}, our resonator achieves this condition with $n \approx 30$ photons in the resonator.
With these, we obtain $\Delta\omega_\text{K} \approx 2\kappa_s$.
The black vertical bar in Fig. \ref{fig:nonlinChar} (e) indicates this shift and matches well with the frequency shift between CB and CD.
}
{
\begin{figure*}[!htb]
    \centering
    \includegraphics[width=\textwidth]{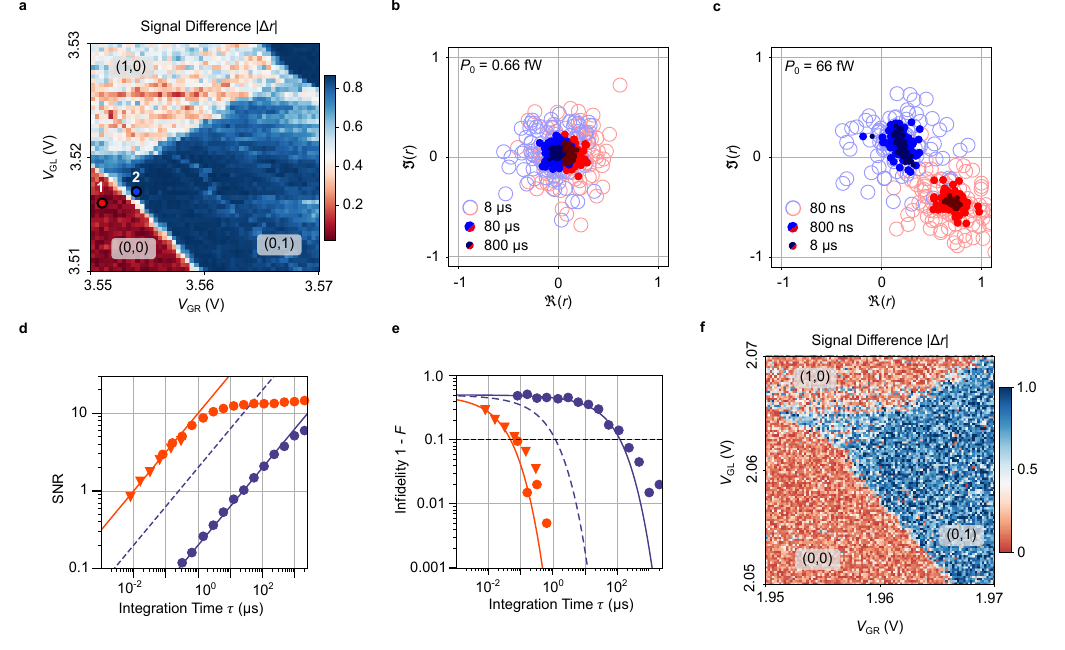}
    \caption{\textbf{Fidelity and signal-to-noise ratio.}
    \textbf{a} 
    Charge stability diagram of the DQD, measured using the nonlinear detector with an integration time $\tau = 2$ ms per point.
    Three of the visible charge configurations on the DQD are labeled as (0,0), (1,0) and (0,1).
    The points 1 and 2 indicate the gate voltages used for the subsequent measurements.
    \textbf{b-c}
    Scatter plot of the detector signal in the linear \& nonlinear response regime with the DQD tuned to point 1 (red data) and point 2 (blue data).
    \textbf{d-e} 
    Calculated signal-to-noise ratio (SNR) and fidelity as a function of the integration time $\tau$ for the linear (purple circles) and nonlinear regime (orange circles).
    For the orange triangles the room temperature setup was set for a bandwidth of 200 MHz while the other data was measured with a bandwidth of 20 MHz.    
    The lines are fits to the data as described in the main text.
    The horizontal dashed line indicates a fidelity of 0.9.
    \textbf{f} A charge stability diagram measured with an integration time $\tau = 80$ ns per point at lower DQD gate voltages of $V_\text{GR} = 1.96$ V and $V_\text{GL} = 2.06$ V. Therefore this plot presents a measurement at another charge configuration. The electron numbers indicated in the figure are with respect to this other configuration in the bottom left state. The $|\Delta r|$ values in panels a and f are determined with respect to the bottom left corner value.
    }
    \label{fig:timedep}
\end{figure*}

Next, we consider the readout speed for measuring a charge state of a QD system.
To do this we use the input power $P_0 = $ \SI{66}{\femto \watt} yielding the maximum signal $|\Delta r | = 0.8$ of Fig. \ref{fig:nonlinChar}
(b).
The corresponding response in the complex plane is shown in Fig. \ref{fig:nonlinChar} (f) with qualitatively different response as compared to the linear case in Fig. \ref{fig:dev} (e).
The measured QD system is a double quantum dot (DQD) residing in the same nanowire structure as the detector, see Fig.~\ref{fig:dev}a.
It couples capacitively to the QD sensor via a metallic coupler.
Figure \ref{fig:timedep}a presents the detector response as a function of the  voltages $V_{\mathrm{GR}}$ and $V_{\mathrm{GL}}$ applied to the plunger gate electrodes of the DQD.
The sensor is tuned into CD for the charge state (0,0) of the DQD.
The numbers here indicate additional electrons in the DQD relative to the
electron occupancy of the DQD in the botton left state.
Increasing $V_\mathrm{GR}$ adds an electron to the right quantum dot, making the DQD switch to the state (0,1). 
This turns off the sensor resulting in the near unity increase of $|\Delta r| = 0.8$ in the reflection signal as we see in the figure. The additional triangular region around the triple point of the charge states is likely caused either by a DQD bias voltage offset or a back action effect.

Figures \ref{fig:timedep} (b) and (c) present scatter plots of the detector signal measured repeatedly for the two charge states at the points 1 and 2 indicated in Fig. \ref{fig:timedep}a.
The data in the linear regime of Fig. \ref{fig:timedep} (b) show that an integration time $\tau = 800\ \mu$s is needed to resolve the two charge states.
The non-linear readout in Fig. \ref{fig:timedep} (c) works significantly faster.
The charge state can be resolved even with the smallest integration time of $\tau = 80$ ns.
This speed improvement arises partially from the larger readout signal used with the larger input power $P_0$ and partially from the increased sensitivity of the nonlinear operation.
To distinguish these effects from each other, we consider the fidelity $F$ and signal-to-noise ratio (SNR) presented in Figs. \ref{fig:timedep} (d-e).
Here, the fidelity is determined as the fraction of the measured data points that fall on the correct side of the discrimination line set halfway between the two charge states \cite{Schaller2010, ChenL2023}.
We see that in the linear case (dark-blue data) a high fidelity of 0.9, corresponding to a $\mathrm{SNR} \approx 3$, is achieved for $\tau = 200$ $\mu$s.
In the nonlinear case (bright-red data), the same condition is obtained already at $\tau = 80$ ns.
The solid lines are determined as $\mathrm{SNR}  = \sqrt{\tau/\tau_0}$ and $F = 1 - 1/(1 + e^{\sqrt{\tau/\tau_0}})$, where $\tau_0 = 25\ \mu$s (10 ns) is the integration time required for $\mathrm{SNR}  = 1$ in the linear (nonlinear) case.
The linear case is measured at 100 times lower power corresponding to 10 times lower reflection signal.
To account for this, we plot the linear response curves shifted to the corresponding 10 times higher signal level as the dashed lines in Fig. \ref{fig:timedep} (d)-(e).
This signal difference explains two orders of magnitude of the readout speed difference.
The nonlinear case still outperforms the linear one further by more than an order of magnitude in speed thanks to the dispersive shift.
Finally, in Fig.~\ref{fig:timedep} (f), we illustrate the fast readout by repeating the measurement of Fig. \ref{fig:timedep}a  at a different charge configuration with $\tau = 80$ ns per point.
This measurement shows clearly that both the $(0,0) \leftrightarrow (1,0)$ and $(0,1) \leftrightarrow (1,0)$ transitions appear with good contrast at this measurement speed.
}

\section{Discussion}
{
Increasing the output signal amplitude $A$ would allow for decreasing the fastest readout time $\tau = 80$ ns of Fig. \ref{fig:timedep} further.
A larger $A$ is achieved by designing the device such that the non-linear regime condition $\omega_\text{K} = -\kappa$ takes place at larger input power $P_0$.
According to Eq. \eqref{eq:nonLinShift}, increasing the number of junctions $N$, the onset power $P_{0}$ and correspondingly the readout speed $\tau$ increases proportionally to $N^2$.
The larger drive power $P_0$ also results in a larger voltage amplitude $V$ in the resonator $V = \left(4\kappa_\text{c}\omega_\text{r}Z_\text{r}P_{0}/\kappa^2\right)^{1/2}$ \cite{Haldar2023}.
With the $\tau = 80$ ns readout, we have $V = 80\ \mu$V.
This is comparable to the typical linewidth of QD sensors in the 100 $\mu$eV-range~\cite{Frey2011, Ares2016, Keith2019, Havir2023}. 
When the microwave amplitude exceeds the sensor linewidth, $\kappa_\text{s}$ reduces \cite{Frey2011, Cornia2019, Havir2023} leading to a contrast reduction between CB and CD.
In order to increase the input power, one therefore has to keep the voltage amplitude low by either increasing the linewidth $\kappa$ of the resonator or by reducing the characteristic impedance $Z_\mathrm{r}$ of the device.

Interestingly, our device scheme has a third option in form of an additional protection scheme against the microwave signal $V$:
First, in CB, the usual protection scheme where the QD is non-conducting~\cite{Morello2010, Park2025} is obtained for our device. When the sensor quantum dot is in CB, the electron tunneling and its stochastic fluctuations are suppressed. This lowers the detector back action for this charge state.
In addition, when the device is at the other charge state with the QD conducting, the resonance frequency of the cavity is shifted away from the frequency of the input signal, see the gray circles in Fig.~\ref{fig:nonlinChar} (c) highlighting that the CD case is closer to the value $|r| = 1$ because of the frequency shift. This implies that a larger fraction of the drive signal reflects directly away from the input port instead of entering to the resonator and causing e.g. dissipative heating~\cite{Keith2019}.
Therefore, also the second charge state is protected from the readout signal.
This may therefore protect the QD sensor against the microwave broadening as well as reducing the other back-action effects, such as heating \cite{Keith2019}, sensor-induced dephasing \cite{Haldar2025} and photo-assisted tunneling \cite{Mavalankar2016}.
Studying the effectiveness of this protection scheme is beyond the scope of this study, and to be investigated further in future works.

As our detector scheme does not need to follow the matching requirement, it opens up new possibilities to design the charge detector.
The response time $\tau_\text{QD}$ of the QD charge detector is set by the $RC_\Sigma$ time constant where $R = 1/G$ and $C_\Sigma$ is the total capacitance of the quantum dot.
Our device has $G = 0.4\ \mu$S and $C_\Sigma = 50$ aF based on the charging energy $E_\text{ch} = e^2/2C_\Sigma = 1.5$ meV seen in Fig. \ref{fig:dev} (c).
This yields an extremely fast response rate of $1/\tau_\text{QD} = G/C_\text{QD} = 8$ GHz, typical for QDs.
However, in the linear response regime, the much slower coupling rate $\kappa_\text{s} = G/C_\text{r}$ needs to be matched with the resonator coupling $\kappa_\text{c}$. 
The overall response time of the detection speed is thus limited to the resonator linewidth taking place at this same slower timescale.
The resonator total capacitance $C_\mathrm{r}$ is $\sim$3 orders of magnitude larger than $C_\Sigma$, slowing down the response time by the same fraction, typical for the various charge detection realizations \cite{Buehler2004, Bylander2005, Angus2005, Dehollain2014, Lu2003, Elzerman2004, Reilly2007, Barthel2010}.
Since the non-linear detection removes the matching requirement, a much larger $\kappa_\text{c}$ can be used without suppressing the signal.
This enables to decrease the resonator response timescale $1/\kappa$, bringing it closer to the response time of the quantum dot $\tau_\text{QD}$ and increasing the response speed correspondingly.
In the numerical calculations shown in Fig. \ref{fig:NumericalCalcs}, we increase $\kappa_\text{c}$ such that $\kappa/\kappa_\text{s} = 10$, while still maintaining the near-unity signal without entering the bifurcation regime.
We therefore predict that at least an order of magnitude increase in the resonator response speed is achievable. 
However, as $\kappa_\text{c}/\kappa_\text{s}$ increases, the frequency shift $\Delta\omega_\text{K}$ becomes small compared to the linewidth.
The decreasing frequency shift eventually leads to a decrease of the reflection signal $|\Delta r|$, setting a limit to $\kappa_\text{c}/\kappa_{s}$.
In addition, the nonlinearities may decrease the response speed of the resonator, see Sect. D in the Supplemental materials for details. Further experiments are therefore needed for testing the achievable speed increase.
}

{
To conclude, we showed that a nonlinear resonator, turns the dissipative response of a charge sensor into a frequency shift of the readout resonator.
This effect allows a near unity signal strength for charge readout without having to satisfy the usual matching requirement in the resonator.
As a result, we achieved an order of magnitude increase in the charge readout speed.
Avoiding the matching requirement opens up an additional protection scheme against back-action effects and to make faster charge detectors approaching RC time constant of the QD sensor in the 1 ns regime.
}

{\bf Acknowledgements} We thank Claes Thelander for fruitful discussions and useful feedback for the manuscript and the European Union (ERC, QPHOTON, 101087343), Swedish Research Council (Dnr 2019-04111 and 2024-04718), the Knut and Alice Wallenberg Foundation through the Wallenberg Center for Quantum Technology (WACQT), and NanoLund for financial support. Views and opinions expressed are however those of the author(s) only and do not necessarily reflect those of the European Union or the European Research Council Executive Agency. Neither the European Union nor the granting authority can be held responsible for them.

\section{Methods}

\subsection{Device Fabrication}
The device is made with 6 lithography steps on top of a high-resistivity $\rho >$ \SI{10}{\kilo \ohm} Si wafer with 200 nm thermally grown oxide. The lithography steps are:
\begin{enumerate}
    \item EBL defined alignment markers and contact pads made with lift-off process and e-beam evaporation of 5 nm Ti + 45 nm Au.
    \item UVL defined feed-lines for DC and RF signals made with lift-off process and e-beam evaporation of 5 nm Ti and 95 nm Al.
    \item  UVL defined shunting capacitance made with a lift-off process. First a 30 nm thick AlOx layer was grown and then a 100 nm thick Al film e-beam evaporated. Following this process step, nanowires were manually transferred, and the barriers located as in Ref. \citealp{Barker2022}. 
    \item  EBL defined window over the nanowire for etching the GaSb shell, Etching for 5 minutes in 2.38\% TMAH. 
    \item  EBL defined nanowire contacts with lift-off process. After development, the native oxide of InAs was removed with 1:10 HF BOE. Etch time was 5 s followerd by rinsing in water and loading immediately to the evaporator. Then 5 nm Ti and 135 nm Al was deposited with e-beam evaporator. 
    \item  EBL defined Niemeyer–Dolan bridge for the resonator array. The resist stack consisted of 3 layers of MMA(8.5)MAA copolymer, 9\% solids in Ethyl Lactate, spin coated with 4000 rpm. The top layer is 950 000 molecular weight PMMA, 6\% solids in Anisole, spin coated with 5000 rpm. The Josephson junctions and the rest of the resonator structure was formed with a two-angle evaporation of Al at $\pm 30^\circ$, with an intermediate oxidation step determining the resistance of the resulting Josephson junctions.
\end{enumerate}

The device was then wire bonded to the measurement PCB and measured at room temperature to determine the resistance of the Josephson junctions and quantum dots.

\subsection{Measurement Setup}
The measurements were performed in a dilution refrigerator at a base temperature of 10 mK. The measurement setup was identical to Ref.~\citealp{Haldar2024} with the following differences: The 80 MHz low pass filter after the mixer was removed, and an amplifier with 200 MHz bandwidth (Femto HVA-200M-40-B with AC coupling and 20 dB gain) was used to allow for a larger bandwidth in the measurement. The signal was then digitized with an oscilloscope and the in-phase and quadrature components were determined. An intermediate frequency (IF) of 12.5 MHz was used for the data presented in Figs.~\ref{fig:dev}, \ref{fig:nonlinChar}, and the 20 MHz data of Fig.~\ref{fig:timedep}. The oscilloscope sampling rate was set to 50 MS/s and the analog bandwidth to 20 MHz. For the 200 MHz data in Fig.~\ref{fig:timedep}, the intermediate frequency was increased 125 MHz, the sampling rate to 500 MS/s with full analog bandwidth of the oscilloscope. Then the IF amplifier Femto HVA-200M-40-B was defining the overall analog bandwidth of the setup to 200 MHz. Note that the intermediate frequencies here were chosen to be approximately a factor of two lower than the analog bandwidth. This lowers the time resolution by a corresponding amount as one needs to have at least one oscillation period of the IF signal in order to determine the response. Using this lower intermediate frequency was done in order to avoid excessive signal distortions such as signal compression and stronger frequency dependence in the attenuation that are induced if the IF signal would have been precisely at the bandwidth specifications of the analog components.


\section{Data Availability}
The data generated in this study have been deposited in the Zenodo database under accession code \hyperlink{15672394}{https://zenodo.org/records/15672394}



\bibliography{master}



\renewcommand{\thefigure}{S\arabic{figure}}
\renewcommand{\theHfigure}{S\arabic{figure}}

\setcounter{figure}{0}

\title{Supplementary Materials to Near-Unity Charge Readout Signal in a Nonlinear Resonator without Matching the Sensor Dissipation}

\author{Harald Havir \orcid{0000-0002-2944-986X}}
\affiliation{NanoLund and Solid State Physics, Lund University, Box 118, 22100 Lund, Sweden}

\author{Andrea Cicovic}
\affiliation{NanoLund and Solid State Physics, Lund University, Box 118, 22100 Lund, Sweden}

\author{Pierre Glidic}
\affiliation{NanoLund and Solid State Physics, Lund University, Box 118, 22100 Lund, Sweden}
\affiliation{Université Lyon 1, CNRS, Institut Lumière Matière, UMR5306, F-69100, Villeurbanne, France}
\author{Subhomoy Haldar}
\affiliation{NanoLund and Solid State Physics, Lund University, Box 118, 22100 Lund, Sweden}
\affiliation{Department of Physics, Indian Institute of Technology Kanpur, Uttar Pradesh 208016, India}

\author{Sebastian Lehmann}
\affiliation{NanoLund and Solid State Physics, Lund University, Box 118, 22100 Lund, Sweden}

\author{Kimberly A. Dick}
\affiliation{NanoLund and Solid State Physics, Lund University, Box 118, 22100 Lund, Sweden}
 \affiliation{Center for Analysis and Synthesis, Lund University, Box 124, 22100 Lund, Sweden}
\author{Ville F. Maisi \orcid{0000-0003-4723-7091}}
\email{ville.maisi@ftf.lth.se}
\affiliation{NanoLund and Solid State Physics, Lund University, Box 118, 22100 Lund, Sweden}

\maketitle

\section{Supplementary Materials}

\subsection{QD sensor characterization}
\begin{figure}
    \centering
    \includegraphics[width = \linewidth]{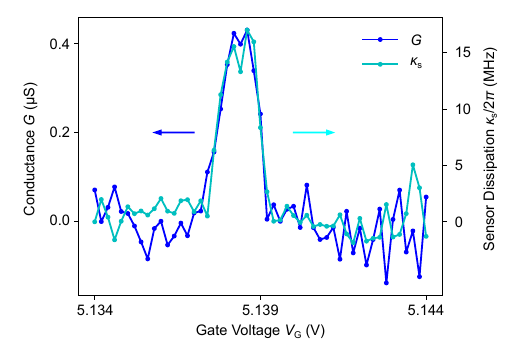}
    \caption{\textbf{Quantum Dot Conductance and Dissipation.} The sensor quantum dot conductance $G$ and dissipation $\kappa_\mathrm{s}$ as a function of the gate voltage $V_\mathrm{G}$.}
    \label{fig:Gandkappas}
\end{figure}

Supplementary Figure~\ref{fig:Gandkappas} presents the sensor QD conductance $G$ and the sensor dissipation $\kappa_\mathrm{s}$ as a function of the gate voltage $V_\mathrm{G}$ for the operation point studied in Figs.~1 - 3a of the main article. The conductance was determined from the slope of the IV curve over a $20\ \mu$V bias range around the zero bias voltage. The sensor dissipation $\kappa_\mathrm{s}$ was determined with the same procedure as for the red data set in Fig. 1d.

\begin{figure}
    \centering
    \includegraphics[width=\linewidth]{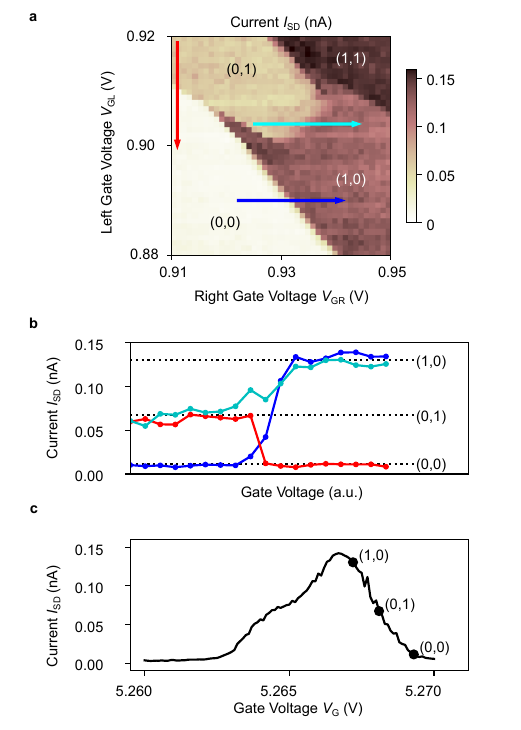}
    \caption{\textbf{Sensor coupling to DQD charge states.} \textbf{a} The sensor quantum dot current $I_\text{SD}$ as a function of the double dot gate voltage $V_\text{GL}$ and $V_\text{GR}$ for one more charge configuration. Note that the DQD gate voltages here differ from the settings used in the main article. The indicated charge occupations $(0,0)$, $(0,1)$, $(1,0)$, $(1,1)$ here indicate again the number of added electrons in relation to the state in the bottom left corner. The QD voltage bias is $V_b = 0.4$ mV. \textbf{b} Linecuts made as indicated in panel \textbf{a}. \textbf{c} The QD sensor response as a function of the gate voltage $V_\mathrm{G}$ for the applied bias voltage $V_b = 0.4$ mV. The dots indicate the three charge states of panel \textbf{a}.}
    \label{fig:sensitivity}
\end{figure}

Supplementary Figure~\ref{fig:sensitivity} presents characterization data that we use for determining the capacitive coupling strength of the DQD to the charge sensor QD. Supplementary Figure~\ref{fig:sensitivity} presents a charge stability diagram of the DQD measured by probing the direct current $I_\text{SD}$ of the sensor QD with a fixed voltage bias $V_b = 0.4$ mV. This method with a finite bias and measuring the direct current $I_\text{SD}$ is chosen here to have a wider gate voltage span with charge sensitivity, see panel \textbf{c} where the sensitivity range with the corresponding charge states $(0,0)$, $(0,1)$ and $(1,0)$ are indicated, and cf. to Supplementary Fig.~\ref{fig:Gandkappas}. The measurement here was made so that the QD sensor was first tuned to the point indicated with "$(0,0)$" in panel c while the DQD gate voltages were set to $V_\text{GL} =  0.91$~V and $V_\text{GR} =  0.88$~V. Then the measurement of panel \textbf{a} was made where these gate voltages are swept. When the DQD occupancy changes, the current $I_\text{SD}$ increases as indicated in panel \textbf{c} for the three charge states $(0,0)$, $(0,1)$ and $(1,0)$. Supplementary Fig.~\ref{fig:sensitivity} presents linecuts along the lines indicated in panel \textbf{a}. These line cuts show the occupancy changes $(0,0) \leftrightarrow (0,1)$, $(0,0) \leftrightarrow (1,0)$ and $(0,1) \leftrightarrow (1,0)$.

To quantify the capacitive coupling between the DQD and the QD sensor, we determine how large fraction of the added electron "gates" the QD sensor. If the capacitive coupling would be very strong, the full charge of the added electron, e.g. for the  $(0,0) \rightarrow (0,1)$ transition, would appear on the sensor QD, and the sensor QD would be shifted by a full Coulomb oscillation period in gate voltage $V_\mathrm{G}$. The capacitive coupling is however much weaker, and as we can see from Supplementary Fig.~\ref{fig:sensitivity}\textbf{c}, the shift in gate voltage is $\Delta V_\mathrm{G} = 1.1$~mV between the charge states $(0,0)$ and $(0,1)$. As one Coulomb oscillation period is $\Delta V_\mathrm{G} = 13$~mV, the $(0,0) \rightarrow (0,1)$  transition induces a $2~\text{mV}/(13~\text{mV/e}) = 0.08$e charge change at the sensor QD. Similarly for the $(0,0) \rightarrow (1,0)$ transition we obtain $\Delta V_\mathrm{G} = 2$~mV leading to $0.15$e, and for the $(0,1) \rightarrow (1,0)$ transition $\Delta V_\mathrm{G} = 0.9$~mV leading to $0.07$e.

\subsection{Circuit Model}
\label{TheoryModel}

For modeling the detector response, we solve the voltages and currents in the circuit of Fig.~1b under the sinusoidal input signal used in the experiment. 
The RF sinusoidal input signal with power $P_0$ and amplitude $A_0$ entering via a $Z_0 = $ \SI{50}{\ohm} transmission line is equivalent of a voltage supply with voltage $V_\mathrm{in}(t) = V_0 \left( e^{i\omega t} + e^{-i\omega t}\right)/2$, and internal resistance of $Z_0$, where $V_0 = |A_0|$. 
The amplitude $V_0$ connects to the input power as $P_0 = (V_0/2\sqrt{2})^2/Z_0$. 
Here the $\sqrt{2}$ accounts for the difference between the RMS value of the voltage and its amplitude, and the factor of two for considering the power $P_0$ at the output after the internal resistance. Here it is important to make sure that the input drive $V_\mathrm{in}(t)$ is properly real-valued. Using e.g. $V_\mathrm{in}(t) = V_0 e^{i\omega t}$, which is often done for linear circuits does not work for the non-linear case. This because taking a real value of this complex-valued solution is not anymore a solution for the non-linear equation. That procedure works only for a linear differential equation where taking the real value yields directly a real-valued solution.

For the Josephson junctions, we use the semiclassical Josephson relations
\begin{equation}
    \label{eq:JJrel}
    \left\{
    \begin{aligned}
        I &= I_0 \sin{\phi} \\
        V &= \frac{\hbar N}{2e}\frac{\partial\phi}{\partial t}, 
    \end{aligned}
    \right.
\end{equation}
for the current $I$ and voltage $V$ across $N$ identical junctions in series.
The Kirchhoff's rules for the components in Fig. 1b 
(conservation of current in the nodes, and voltages adding up to zero in the loops), yield then the equation of motion
\begin{equation}
    \label{eq:eom}
    \frac{\partial^2\phi}{\partial t^2} + (\kappa_\text{s} + \kappa_\text{c}) \frac{\partial \phi}{\partial t} + \omega_\text{r}^2 \sin{\phi} = c_\text{c}\omega_\text{r} \frac{\partial v_\mathrm{in}(t)}{\partial t},
\end{equation}
for the phase $\phi$ across the array. 
Here $\omega_\text{r} = 1/\sqrt{L C_\text{r}}$ is the resonance frequency in the low power linear regime with inductance $L = \frac{\hbar N}{2e I_0}$, and capacitance $C_\text{r} = C + C_\text{c}\left(1+Z_0G\right)$, and $c_\text{c} = C_c/C_\text{r}$. The QD loss term is $\kappa_\text{s} = G/C_\text{r}$, i.e. the RC time constant arising from the conductance $G$.
Note here that the finite frequency conductance usually differs from the low-frequency one \cite{Havir2023}, leading to a difference between the low-frequency and high-frequency dissipation.
The loss via the input port takes the usual~\cite{Goppl2008, Andersson2025} form $\kappa_\text{c} = \frac{Z_0 C_1^2 \omega_\text{r}^2}{C_\text{r}} \left(1+Z_0G\right)$, when combining the terms $Z_0 C_1 \omega_\text{r}^2 \cos \phi\, \frac{\partial \phi}{\partial t}$ and $\frac{Z_0 C C_\text{c}}{C_\text{r}}  \frac{\partial^3 \phi}{\partial t^3}$. 
Here we approximated $\cos \phi = 1$ to the leading order in $\phi$, and consider a solution of the form $\phi = \phi_1 e^{i\omega t} + \phi_1^* e^{-i\omega t}$. 
This ansatz allows to approximate $\frac{\partial^3 \phi}{\partial t^3} = -\omega^2 \frac{\partial \phi}{\partial t} = -\omega_\text{r}^2 \frac{\partial \phi}{\partial t}$ close to the resonator frequency $\omega_\text{r}$ to obtain the above result for $\kappa_\text{c}$. 
The right hand side of Eq.~(\ref{eq:eom}) is the drive term with normalization $v_\mathrm{in}(t) = \frac{2e}{N \hbar \omega_\text{r}} V_\mathrm{in}(t)$.

Next with the Taylor expansion \mbox{$\sin{\phi}\approx \phi-\phi^3/6$}, Eq.~(\ref{eq:eom}) has the form of a Duffing oscillator. 
The harmonic balance method yields then the equation
\begin{equation}
    \label{eq:phi1}
    \phi_1 = \frac{i\omega c_\text{c} v_0 / 2\omega_\text{r}}{1 - \omega^2/\omega_\text{r}^2 + i\omega\kappa/\omega_\text{r}^2 - |\phi_1|^2/2},
\end{equation}
for the amplitude $\phi_1$ of the phase response when solving the equation for the fundamental mode at $\omega$ and neglecting the higher frequency components. 
Equation~(\ref{eq:phi1}) is easily solved with fixed point iteration to obtain the amplitude $\phi_1$. 
Here $v_0 = \frac{2e}{N \hbar \omega_\text{r}} V_0$, is the normalized input amplitude. 
Then the resonator impedance $Z$ is determined by calculating the voltage and current in front of the input capacitor $C_\text{c}$. 
This results in
\begin{equation}
    Z = \frac{v_0/2 - \omega^2 C_\text{c} Z_0 \phi_1/\omega_\text{r}}{i\omega C_\text{c}(v_0/2 - i\omega \phi_1/\omega_\text{r})}.
\end{equation}
The reflected signal is then finally given by the standard formula
\begin{equation}
    r = \frac{Z-Z_0}{Z+Z_0}.
\end{equation}
The reflection coefficient $r$ here is often denoted with $\Gamma$ instead of the $r$. We chose to use $r$ to avoid confusion with the usual notation in our field where $\Gamma$ is often used to indicate various rates.

Supplementeray Figure~\ref{fig:ExpAndNum} presents the experimental data for both CB and CD cases and their difference on the left column. The corresponding circuit model calculations are given on the right column.
{The parameter values of the device are summarized in Supplementary Table \ref{tab:res_params}. These are used for all the modeling.
\begin{figure}
    \centering
    \includegraphics[width=\linewidth]{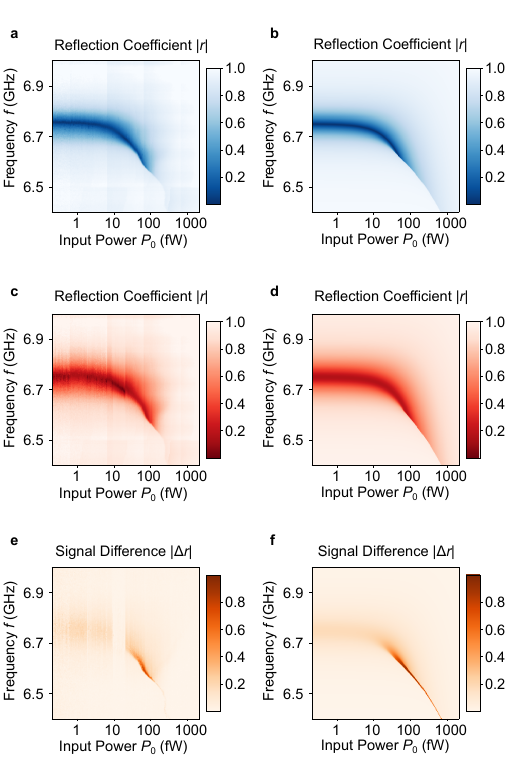}
    \caption{\textbf{Experiment and Simulation of Nonlinear Response.} \textbf{a} Measured resonator response $|r|$ for CB as in Fig. 2 \textbf{a} of the main manuscript. \textbf{b} Corresponding circuit model result.   
    \textbf{c} \& \textbf{d} Same data as in panels \textbf{a} and \textbf{b} but now in CD.
    \textbf{e} \& \textbf{f} the resulting signal difference $|\Delta r|$ between the CD and CB results.}
    \label{fig:ExpAndNum}
\end{figure}

\begin{table}[t]
    \caption{\textbf{Summary of the parameters and their values.}}
    \label{tab:res_params}
    \centering
    \begin{tabular}{lll}
    \toprule
    Name & Parameter & Value \\
    \hline
        Input coupling rate & $\kappa_\mathrm{c}/2\pi$ & \SI{62}{\mega \hertz}  \\
        Internal losses & $\kappa_\mathrm{i}/2\pi$ & \SI{60}{\mega \hertz} \\
        Sensor dissipation rate & $\kappa_\mathrm{s}/2\pi$ & \SI{28}{\mega \hertz} \\
        Resonance frequency & $\omega_\mathrm{r}/2\pi$ & \SI{6.75}{\giga \hertz} \\
        Number of junctions& $N$ & 13 \\
        Junction resistance & $R_J$ & \SI{1.5}{\kilo \ohm}\\
        Junction inductance & $L_J$ & \SI{1.7}{\nano \henry} \\
        Resonator total capacitance & $C_r$ & \SI{25}{\femto \farad} \\
        Coupler capacitance & $C_c$ & \SI{10}{\femto \farad} \\
        Resonator impedance & $Z_\mathrm{r}$ & \SI{950}{\ohm} \\
        \hline
    \end{tabular}

\end{table}

\subsection{Estimation of the resonator response speed reduction from nonlinear effects}
The response time considerations of the main article are based on the response time of a linear system. The nonlinear effects may change this response time and possibly slow down the detection. We estimate this reduction by estimating how much the coupling term $\kappa_\mathrm{c}$ changes for the operation point used in Fig.~3. The circuit model yields the highest phase amplitude of $|\phi_1| =  0.1 \pi$ for this considered operation point. The nonlinear effects contribute therefore at most $5\ \%$ corrections to the linear leading order terms of the equation of motion Supplementary Equation~(\ref{eq:eom}) via the $\sin \phi$ and $\cos \phi$ terms. The reduction of $\kappa_\mathrm{c}$  arises partly from the reduction of the used frequency $\omega$ and partly from the correction to the approximation $\cos \phi \approx 1$ made in the derivation of $\kappa_\mathrm{c}$. Using the observed frequency shift, and the highest phase value of $\phi =|\phi_1| =  0.1 \pi$ to estimate the error in the above approximation, reduces the $\kappa_\mathrm{c}$ value by $5\ \%$. We estimate that for the considered operation point, the reduction of the response time is of similar size, and hence does not slow down the detector significantly. This assessment is further supported by Refs.~\citealp{Bhupathi2016, Johannessen2017, ChenQ2023, Moatimid2023} reporting the transient response in the nonlinear regime taking place in a similar timescale as in the linear response regime. Further experiments probing the response time are, however, needed for testing if this estimation is valid.}

\subsection{The Kerr coefficient}
\label{app:KerrTerm}
The Kerr term with the coefficient $E_\text{K}$ describes the non-linearity of a resonator. 
With the resonance frequency $\omega_\text{r}$ and photon creation and annihilation operators $\hat{a}^\dagger$ and $\hat{a}$, the corresponding Hamiltonian reads
\begin{equation}
    \hat{H} = \hbar \omega_\text{r} \hat{a}^\dagger \hat{a} + \frac{E_\text{K}}{2}\hat{a}^\dagger\hat{a}^\dagger\hat{a} \hat{a},
\end{equation}
resulting in the energy difference between $n$ and $n-1$ photons as
\begin{equation}
    E_n = \hbar \omega_\text{r} + E_\text{K} (n-1).
\end{equation}
In other words, the resonance frequency is shifted by $\omega_\text{K} = \frac{E_\text{K}}{\hbar}(n-1)$ for $n$ photons in the resonator, relative to the lowest photon transition frequency of $E_1/\hbar = \omega_\text{r}$. 
For the weak non-linearity considered in this work, this shift is only significant for $n \gg 1$ allowing us to approximate $n-1 \approx n$ for the shift.

The Kerr coefficient is $E_\text{K} = -E_\text{C}$ for a single Josephson junction~\cite{Koch2007, Kirchmair2013, Hoi2013, Yamaji2022, Andersson2025}. 
For the $N$ junction array considered in this work, the total resonator voltage $V$ is divided across the $N$ junctions. 
The photon number $n$ in the resonator is proportional to the energy stored in the resonator, which in turn is proportional to $V^2$. 
Therefore, to reach the same voltage amplitude across a single junction in the array, the photon number needs to be $N^2$ times larger as compared to the single junction case. 
Therefore, the Kerr coefficient for the array is $E_\text{K} = - E_\text{C}/N^2$. 
Here it is important to note that the "charging energy" $E_\text{C} = e^2/2C_\text{r}$ is calculated for the resonator total capacitance $C_\text{r}$. 
It is therefore the charging energy of the resonator, not that of the QD. 
Using the relation $\omega_\text{r} = 1/\sqrt{L C_\text{r}}$, for the resonance frequency and $Z_\text{r} = \sqrt{L/C_\text{r}}$, for the characteristic impedance of the resonator at low power limit, the Kerr coefficient becomes
\begin{equation}
    \label{eq:kerrCoef}
    E_\text{K} = -\hbar \omega_\text{r} \frac{\pi Z_\text{r}}{R_\text{Q} N^2},
\end{equation}
where $R_\text{Q} = h/e^2$, is the resistance quantum. 
We therefore see that the characteristic impedance $Z_\text{r}$ and the number of junctions $N$ are the key parameters determining the resonator non-linearity via the Kerr coefficient.

Considering further that the relation between the input power $P_0$ and the number of photons $n$ in the linear resonator response regime~\cite{Haldar2023, Andersson2025} is
\begin{equation}
    \label{eq:nphot}
    n = \frac{4 \kappa_\text{c}}{\kappa^2} \frac{P_0}{\hbar \omega_\text{r}},
\end{equation}
we obtain the frequency shift as
\begin{equation}
    \label{eq:omegaK}
    \omega_\text{K}/\omega_\text{r} = -\frac{4\pi Z_\text{r} \kappa_\text{c}}{R_\text{Q} N^2 \kappa^2}   \frac{P_0}{\hbar \omega_\text{r}}.
\end{equation}
The non-linearity becomes visible in the response when it is comparable to the resonator linewidth: $\omega_\text{K} = -\kappa$. 
This condition yields the threshold input power between the linear and non-linear response regime as
\begin{equation}
    \label{eq:Pth}
    P_{0} = \frac{\hbar R_\text{Q} N^2 \kappa^3}{4\pi Z_\text{r} \kappa_\text{c}}.
\end{equation}
The frequency shift $\Delta \omega_\text{K} = 2\kappa_\text{s}$ resulting in from the sensor dissipation $\kappa_\mathrm{s}$ is obtained by using this power and Taylor expanding Supplementary Equation~\eqref{eq:omegaK} in $\kappa_\text{s}$.

\subsection{Bifurcation threshold}
This section shows that the optimal input power for the nonlinear regime is indeed obtained at the $\omega_K \approx -\kappa$ condition. We consider the bifurcation threshold condition~\cite{Ong2011}, that is, the largest input power $P_0$ where the resonator still has a unique amplitude solution $\left| \phi_1 \right|$ for all input frequencies. We also show that the above relation as well as Supplementary Equation~(\ref{eq:Pth}) - that was obtained with the linear resonator photon number of Supplementary Equation~(\ref{eq:nphot}) - is correct within $\sim 20~\%$, and obtain an analytical equation for the amplitude $\left| \phi_1 \right|$ at the bifurcation threshold.

\begin{figure}
    \centering
    \includegraphics[width=\linewidth]{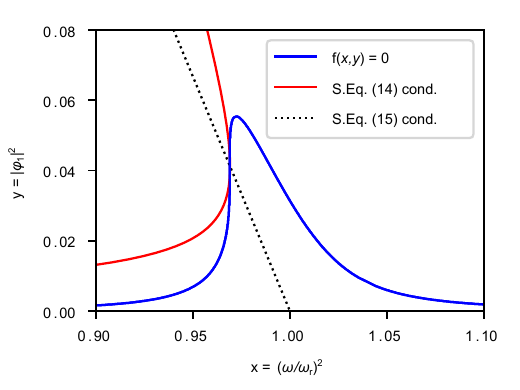}
    \caption{\textbf{Resonator response at the bifurcation threshold.} The solid blue line shows the resonator phase amplitude $\left| \phi_1 \right|$ as a function of the input drive frequency $\omega$ for the studied resonator with $\alpha = \left(\kappa/\omega_\mathrm{r} \right)^2 = 3.3 \cdot 10^{-4}$, and input drive value at the bifurcation threshold value of $\beta = 16 \alpha^{3/2}/3\sqrt{3} = 1.8 \cdot 10^{-5}$. The curve is obtained by solving numerically the condition $f(x,y)=0$. The solid red curve plots Supplementary Eq.~(\ref{eq:gxy}) and the dashed black line Supplementary Eq.~(\ref{eq:dgdy}). The three curves cross at the bifurcation onset point $x = 1 -\sqrt{3 \alpha} = 0.97,\ y =4\sqrt{\alpha/3} = $0.042, with $\partial f/\partial y = \partial g/\partial y = 0$ as required for the onset point determining the bifurcation threshold.}
    \label{fig:DuffSol}
\end{figure}

We start by rewriting Supplementary Equation~(\ref{eq:phi1}) as
\begin{equation}
    \label{eq:phi1sq}
    \left[\left(1-x-y/2\right)^2 + \alpha x \right] y = \beta x,
\end{equation}
where $x = \left(\omega/\omega_\mathrm{r} \right)^2$, $y = \left| \phi_1 \right|^2$, $\alpha = \left(\kappa/\omega_\mathrm{r} \right)^2$ and $\beta = \left(c_c v_0/2\right)^2$. For the coming argumentation, it is useful to define the function
\begin{equation}
    \label{eq:fxy}
    f(x,y) = \left[\left(1-x-y/2\right)^2 + \alpha x \right] y - \beta x.
\end{equation}
The solutions of Supplementary Equation~(\ref{eq:phi1sq}) are then $f\left(x,y \right)=0$, yielding the possible values of the oscillation amplitudes $y$ for a given input frequency $x$. Supplementary Figure~\ref{fig:DuffSol} plots these solutions as a solid blue line with the value $\alpha = 3.3 \cdot 10^{-4}$ valid for the resonator in the main manuscript, and input drive of $\beta = 1.8 \cdot 10^{-5}$.

The bifurcation regime takes place for large input drive $\beta = \left(c_c v_0/2\right)^2$ such that the $\omega < \omega_\mathrm{r}$ side has multiple solutions for $\left| \phi_1 \right|$. The bifurcation threshold, i.e. drive value $\beta$ above which multiple solutions exist, needs to fulfill {\it Condition I:} The blue curve satisfying $f(x,y) = 0$ in Supplementary Figure~\ref{fig:DuffSol} has one and only one point for which $dy/dx \rightarrow \infty$. This condition is fulfilled if and only if {\it Condition II:} $\partial f(x,y)/\partial y = 0$, and {\it Condition III:} The $\partial f(x,y)/\partial y = 0$ curve is tangential to the $f(x,y)=0$ curve, are both satisfied at that one point. The Condition II makes sure that $dy/dx \rightarrow \infty$ is satisfied, and the Condition III with the two curves tangential makes sure that there is only one point where Condition II is satisfied. If Condition III is not fulfilled, either no solutions exists with the curves never crossing (small $\beta$) or two solutions exist as two crossing points (large $\beta$ in the bifurcation regime). The values of $x$, $y$ and $\beta$ at that single point describe the bifurcation threshold. The multivalued amplitudes $\left| \phi_1 \right|$ develop around this point for the larger input drives. Therefore, it makes sense to call this point the bifurcation threshold point.

The Condition II is easily determined to yield the equation
\begin{equation}
    \label{eq:gxy}
    g(x,y) = \left(1-x-y/2\right) y^2 - \beta x = 0.
\end{equation}
Here we have evaluated the partial derivative $\partial f/\partial y = 0$, used the fact that $f(x,y) = 0$ for the considered point, and multiplied the resulting equation by $y$. 
The solution of Supplementary Equation~(\ref{eq:gxy}) is shown as solid red line in Supplementary Figure~\ref{fig:DuffSol}. The Condition III, $\partial f(x,y)/\partial y=0$ tangential to $f(x,y)=0$, is equivalent to $\partial g/\partial y = 0$, i.e. that $g(x,y) = 0$ is tangential to $f(x,y)=0$, when noting that the tangent of $f(x,y) = 0$ is along the y-axis for the bifurcation threshold point. 
With the requirement $\partial g/\partial y = 0$, we obtain thus
\begin{equation}
    \label{eq:dgdy}
    1 - x = \frac{3}{4}y,
\end{equation}
which is presented at dashed black line in Supplementary Figure~\ref{fig:DuffSol}. Substituting Supplementary Equation~(\ref{eq:dgdy}) to Supplementary Equation~(\ref{eq:gxy}) yields $\beta x = y^3/4$. Substituting this and Supplementary Equation~(\ref{eq:dgdy}) further to the condition $f(x,y) = 0$ results in $y/4 = \sqrt{\alpha/3}$. Here we have assumed $\alpha \ll 1$, i.e. that the linewidth is much smaller than the resonance frequency, $\kappa \ll \omega_\mathrm{r}$, a condition typically valid for resonators. 

As a summary, for a given resonator linewidth $\alpha$, the bifurcation threshold takes place at the frequency $x$, amplitude $y$ and input drive $\beta$ given by
\begin{equation}
    \label{eq:bifurTresh_xybeta}
    \left\{ \begin{array}{ccc}
         1-x &=& \sqrt{3 \alpha}  \\
         y &=& 4\sqrt{\alpha/3}  \\
         \beta &=& 16 \alpha^{3/2}/3\sqrt{3}   
    \end{array} \right..
\end{equation}
In the non-normalized units, this bifurcation threshold point is expressed as
\begin{equation}
    \label{eq:bifurTresh_wphiP}
    \left\{ \begin{array}{rcl} 
         \omega-\omega_\mathrm{r} &=& -\displaystyle\frac{\sqrt{3}}{2} \kappa  \\[12pt] 
         \left| \phi_1 \right|^2 &=& \displaystyle\frac{4 \kappa}{\sqrt{3} \omega_\mathrm{r}}  \\[12pt] 
         P_0 &=& \displaystyle\frac{\hbar R_\text{Q} N^2 \kappa^3}{3\sqrt{3}\pi Z_\text{r} \kappa_\text{c}},   
    \end{array} \right..
\end{equation}
where a Taylor expansion $1-x = 1 - \left(\omega/\omega_\mathrm{r} \right)^2 = 2(\omega-\omega_\mathrm{r})/\omega_\mathrm{r} + \mathcal{O}\left(  \left[(\omega-\omega_\mathrm{r})/\omega_\mathrm{r}\right]^2\right) $ around $\omega_\mathrm{r}$ was used for the first expression. We see that the first equation matches with the condition $\omega_K = -\kappa$ and the last one with the condition of Eq.~(\ref{eq:Pth}) with just the prefactors differing by less than $25 \%$. In addition, we obtained a relation for the oscillation amplitude $\left| \phi_1 \right| \propto \sqrt{\kappa}$ at the bifurcation threshold.

\begin{figure}[hbt!]
    \centering
    \includegraphics[width=\linewidth]{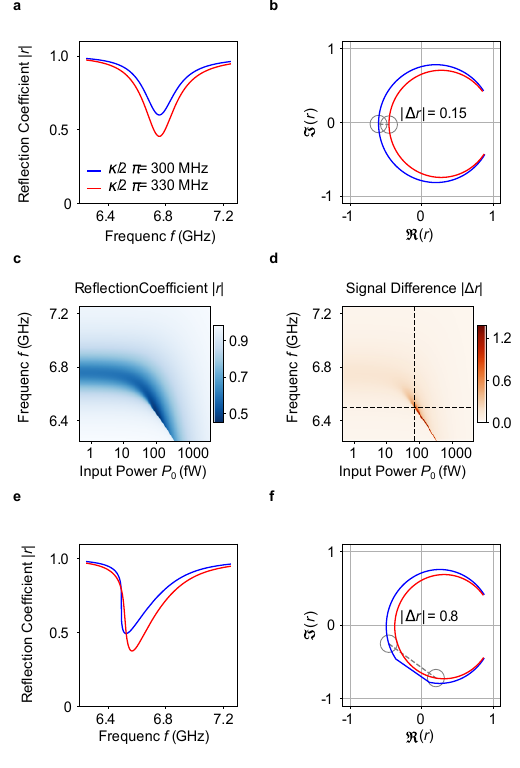}
    \caption{\textbf{Simulated response for larger input/output coupling} Results from simulating a device with $\kappa_\text{c}/2\pi = 240$ MHz, $\kappa_\text{i}/2\pi = 60$ MHz and $\kappa_\text{s}/2\pi = 30$ MHz. 
    \textbf{a} The low-power amplitude response as a function of the frequency $f$. 
    \textbf{b} The corresponding plot of the complex response along with the maximum signal difference $|\Delta r|$.
    \textbf{c} \& \textbf{d} The reflection coefficient $|r|$ \& signal difference $|\Delta r|$ as a function of the frequency $f$ and input power $P_0$.
    \textbf{e} The high-power amplitude response as a function of the frequency $f$, at the power $P_0 = 71 $ fW, indicated by the vertical line in panel \textbf{d}.
    \textbf{f} The corresponding plot of the complex response along with the maximum signal difference $|\Delta r|$ at low power.
    }
    \label{fig:NumericalCalcs}
\end{figure}

\subsection{Simulated Device Data for $\kappa_\mathrm{c} \gg \kappa_\mathrm{s}$}
In Supplementary Figure~\ref{fig:NumericalCalcs} we present simulated device data for a case where the input-output coupling rate $\kappa_\text{c}$ is approximately an order of magnitude larger than the dissipation rate $\kappa_\text{s}$. This is done so that the  dissipation $\kappa_\text{s}/2\pi = 30$ MHz, and internal losses $\kappa_\text{i}/2\pi = 60$ MHz are kept the same as in the measured device, but where the input/output coupling is increased to $\kappa_\text{c}/2\pi = 240$ MHz.
All the other parameter values are the same as for the measured device.
Here the linewidth $\kappa/2\pi = $ 300 MHz is an order of magnitude greater than the sensor dissipation $\kappa_\text{s}$.
Despite being an order of magnitude away from the matching condition, the calculations still predict near-unity signal strength when operating the device at an input power just below the onset of the nonlinear bifurcation.
Here, the onset power where $\Delta \omega_\text{K} = -\kappa$ is $P_0 = 71$ fW, with a corresponding voltage amplitude $V = 60$ $\mu$V. 

With this coupling configuration, and a doubling of the number of junctions in the resonator, one could quadruple the input power without consequence to the sensor contrast.
Additionally, our measurement setup has $\sim$ 3 dB of cable losses between the device and the first amplifier at 4 K, which could be replaced with superconducting cables to increase the signal strength at the amplifier input, granting another factor 2 in speed.
With these considerations, achieving a measurement time of 10 ns is deemed realistic.

This demonstrates the prospects of a better optimized device to reach sub-$10$ ns response time, permitting charge detection well-within the dephasing times of charge- and spin qubits alike.

\end{document}